\documentclass[aps,prl,twocolumn,superscriptaddress]{revtex4}
\usepackage{epsf,graphicx}
\usepackage{amssymb}
\usepackage{amsmath}
\usepackage{amsthm}
\usepackage{latexsym,bm,array,amsfonts,multirow}
\usepackage{color}
\usepackage{ulem}

\begin{document}

\title{Generalized Spectral Decomposition for Quantum Impurity Problems}
\author{Jun-Bin Wang}
\affiliation{Beijing National Laboratory for Condensed Matter Physics and Institute of
	Physics, Chinese Academy of Sciences, Beijing 100190, China}
\affiliation{University of Chinese Academy of Sciences, Beijing 100049, China}
\author{Dongchen Huang}
\affiliation{Beijing National Laboratory for Condensed Matter Physics and Institute of
Physics, Chinese Academy of Sciences, Beijing 100190, China}
\affiliation{University of Chinese Academy of Sciences, Beijing 100049, China}
\author{Yi-feng Yang}
\email[]{yifeng@iphy.ac.cn}
\affiliation{Beijing National Laboratory for Condensed Matter Physics and Institute of
Physics, Chinese Academy of Sciences, Beijing 100190, China}
\affiliation{University of Chinese Academy of Sciences, Beijing 100049, China}
\affiliation{Songshan Lake Materials Laboratory, Dongguan, Guangdong 523808, China}
\date{\today}

\begin{abstract}
Solving quantum impurity problems may advance our understanding of strongly correlated electron physics, but its development in multi-impurity systems has been greatly hindered due to the presence of shared bath. Here, we propose a general operation strategy to disentangle the shared bath into multiple auxiliary baths and relate the problem to a spectral decomposition problem of function matrix for  applying the numerical renormalization group (NRG). We prove exactly that such decomposition is possible for models satisfying (block) circulant symmetry, and show how to construct the auxiliary baths for arbitrary impurity configuration by mapping its graph structure to the subgraph of a regular impurity configuration. We further propose an approximate decomposition algorithm to reduce the number of auxiliary baths and save the computational workload. Our work reveals a deep connection between quantum impurity problems and the graph theory, and provides a general scheme to extend the NRG applications for realistic multi-impurity systems.
\end{abstract}
\maketitle

\begin{figure}[b]
	\begin{center}
		\includegraphics[width=8.6cm]{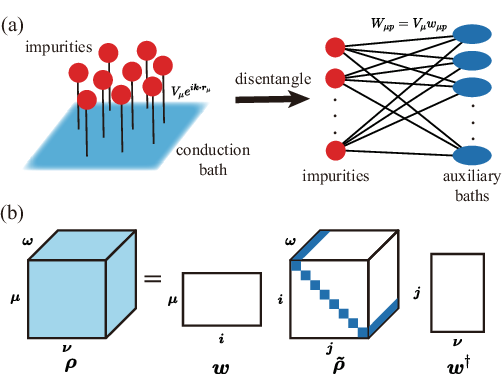}
	\end{center}
	\caption{(a) Disentanglement of the shared bath (left) into multiple auxiliary baths (right), each coupled in a similar momentum-independent form to all impurities. (b) Decomposition of the spectral function matrix $\rho_{\mu\nu}(\omega)$ with the $\omega$-independent transformation matrix $w$ as given in Eq.~(\ref{basic}).}
	\label{fig1}
\end{figure}

\paragraph{Introduction.}
Quantum impurity models play an important role in modern condensed matter physics and practical applications \cite{Cronenwett1988, Glazman2001, Wiel2002, Pustilnik2004, Otte2008, Bork2011,Hiraoka2017, Iftikhar2018,Andersion1961, Kondo1964, Cragg1980, Hewson1997, Vojta2006, Lopes2020}. Numerical renormalization group (NRG) \cite{Wilson1975, Bulla2008} is a powerful tool to solve the quantum impurity models \cite{Bulla2001, Jones1987, Zhu2011}, but its application has been greatly limited by its basic methodology that requires to map the conduction electron bath into a Wilson chain. In reality, multiple impurities often share a common bath with different momentum-dependent couplings that do not allow such a direct mapping. This issue was recently relieved by the so-called auxiliary-bath  approach that disentangles the shared bath into a number of independent auxiliary baths as illustrated in Fig.~\ref{fig1}(a). Each auxiliary bath couples to all impurities in a similar form, and hence can be treated using well-developed traditional NRG algorithms \cite{Hu2023}.

The original and auxiliary-bath models must yield the same inter-impurity correlations after integrating out the baths' degrees of freedom. This requires \cite{Hu2023}
\begin{equation}\label{basic}
\rho_{\mu \nu}(\omega)=\sum_{p} w_{\mu p} w^{*}_{\nu p} \tilde{\rho}_{p}(\omega),
\end{equation}
where $\rho_{\mu \nu}(\omega)=\rho^*_{\nu \mu}(\omega)$ are the (nonlocal) spectral function of the original shared bath between two impurity sites $\bm{r}_\mu$ and $\bm{r}_\nu$ ($\mu,\,\nu=0,\,\cdots,\,N-1$), $\tilde{\rho}_p(\omega)$ are non-negative real functions representing the densities of states of the auxiliary baths ($p=0,\,\cdots,\,N_A-1$), and $w_{\mu p}$ are transformation parameters satisfying $\sum_{p} w_{\mu p} w^{*}_{\nu p} = \delta_{\mu \nu}$. A simple solution with $N_A=4$ has been proposed for 3-impurity models \cite{Hu2023}, but a general strategy has not been established for arbitrary $N$-impurity models. The difficulty comes from the $\omega$ dependency, without which $\rho_{\mu\nu}$ is simply an Hermitian matrix and can be easily diagonalized with $\tilde{\rho}_p$ being its eigenvalues. But it is usually impossible to have the same transformation matrix $w_{\mu p}$ for all $\omega$. We are faced with a generalized spectral decomposition problem as graphed in Fig.~\ref{fig1}(b). 

In this work, we prove that a decomposition with $N_A=N$ can be constructed for special models satisfying the (block) circulant symmetry. By connecting the impurity configuration to a graph structure, we further show that the shared bath of any $N$-impurity model can be decomposed by mapping its graph to the subgraph of a (block) circulant symmetric $N_A$-impurity model, albeit with $N_A>N$. To save the computational workload, we also propose an algorithm to find approximate solutions with reduced $N_A$. These provide a systematic scheme to construct the auxiliary baths and solve the multi-impurity problems using the NRG.

\begin{figure}[t]
	\begin{center}
		\includegraphics[width=8.6cm]{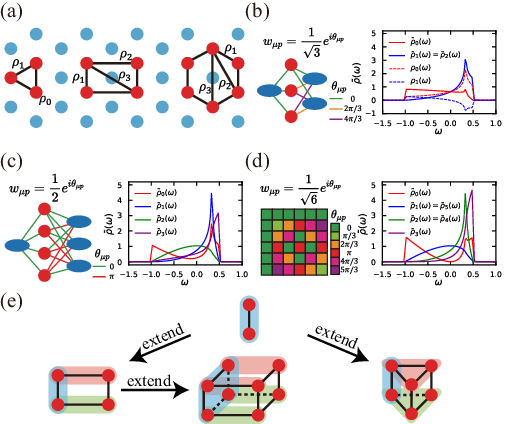}
	\end{center}
	\caption{(a) Examples of regular impurity configurations (regular triangle, rectangle, regular hexagon) on a triangle lattice with the local, nearest, next-nearest, and next-next-nearest-neighbor $\rho_{\mu\nu}(\omega)$ denoted by $\rho_0$, $\rho_1$, $\rho_2$, and $\rho_3$, respectively. (b)(c)(d) Illustrations of the auxiliary-bath models with $N_A = N$ for configurations shown in (a). Also given are their respective transformation parameters $w_{\mu p}$ and densities of states $\tilde{\rho}_p(\omega)$ of the auxiliary baths. For comparison, (b) also shows the local and nearest-neighbor $\rho_{\mu\nu}(\omega)$ of the original bath (the bandwidth is set to 1.5). (e) Extension of a two-impurity configuration into a rectangle, a straight triangular prism, and a cuboid, whose spectral function matrices are block circulant.}
	\label{fig2}
\end{figure}

\paragraph{Exact decomposition.}
Our purpose is to decompose the $N\times N$ spectral function matrix $\rho_{\mu\nu}(\omega)$ into $N_A$ real functions $\tilde{\rho}_p(\omega)$ using an $\omega$-independent transformation matrix $w_{\mu p}$. We first prove that $\rho_{\mu\nu}(\omega)$ satisfying (block) circulant symmetry can be exactly decomposed with $N_A=N$. These correspond to impurity models with highly symmetric real space configurations such as regular triangle, rectangle, or regular hexagon on a triangle lattice as shown in Fig.~\ref{fig2}(a). Our proof contains two steps.

\textbf{Theorem 1.} A regular $N$-gon satisfying the circulant symmetry ($C_N$) can be exactly decomposed with $N_A=N$.

For these models, the spectral function matrix $\rho_{\mu\nu}(\omega)$ has an $n\times n$ circulant form, 
\begin{equation}
\begin{pmatrix}
		a_0 & a_1 & \cdots & a_{n-1}\\
		a_{n-1} & a_0 & \cdots & a_{n-2}\\
		\vdots & \vdots & \ddots &\vdots \\
		a_1 & a_2 & \cdots & a_0
	\end{pmatrix},
\end{equation}
where $\{a_i\}$ are matrix elements depending on $\omega$. It is straightforward to verify that for any values of $a_i$, the above matrix can be diagonalized by the transformation matrix $w^{(n)}_{jk} = e^{i\theta_{jk}}/\sqrt{n}$ with $\theta_{jk}=2\pi jk/n$ for $j,k = 0,\, 1,\, 2,\, \cdots,\, n-1$ \cite{Gray2006}. Thus, for $N$-impurity models satisfying $C_N$ symmetry ($n=N$), the spectral function matrix $\rho_{\mu\nu}(\omega)$ can be decomposed using the same transformation matrix  independent of $\omega$. Regular triangles ($C_3$) and hexagons ($C_6$) on a triangle lattice shown in Fig.~\ref{fig2}(a) are two examples. Figures~\ref{fig2}(b)(d) give the densities of states of their auxiliary baths and the transformation matrices, respectively. $C_4$ symmetric 4-impurity models on a square lattice also belong to this category. 

\textbf{Theorem 2.} All $N\times N$ spectral function matrices satisfying block circulant symmetry can be exactly decomposed with $N_A=N$.

A block circulant matrix takes the form \cite{Tee2007}:
    \begin{equation}\label{eq6}
\begin{pmatrix}
		A_0 & A_1 & \cdots & A_{n-1}\\
		A_{n-1} & A_0 & \cdots & A_{n-2}\\
		\vdots & \vdots & \ddots &\vdots \\
		A_1 & A_2 & \cdots & A_0
	\end{pmatrix},
\end{equation}
where $\{A_i\}$ are $m\times m$ matrices satisfying $A_i = A^{\dagger}_{n-i}$ with $i=0,\, 1,\, \cdots,\, n-1$. If all $A_i$ can be diagonalized by the same transformation matrix $U$, the above matrix can be diagonalized using the direct product $w^{(n)} \otimes U$. This seems quite straightforward. However, theorem 1 tells that for $U$ to be $\omega$-independent, the $A_i$ matrices should also have circulant form. We have therefore $U=w^{(m)}$. By setting $N = nm$, the corresponding $N$-impurity model can be decomposed using the transformation matrix $w= w^{(n)}\otimes w^{(m)}$. This procedure can continue by requiring that $A_i$ are also block circulant. 

The above extension is by no means artificial. The $m$-block may be geometrically understood if each impurity of a regular $n$-gon  represents a cluster of a regular $m$-gon in some perpendicular or internal dimension. The extended impurity model then satisfies the extended $C_n\otimes C_m$ symmetry. Three examples are given in  Fig.~\ref{fig2}(e). A rectangle model may be viewed as an extension of a two-impurity model, where each impurity is extended in a similar manner into two impurities along a perpendicular direction. The model satisfies the $C_2\otimes C_2\cong D_2$ symmetry but can be solved exactly with $N_A=N=4$, despite the lack of $C_4$ symmetry. The corresponding transformation parameters ($w^{(2)}\otimes w^{(2)}$) and the densities of states of the auxiliary baths are given in Fig.~\ref{fig2}(c). Similarly, we may extend each impurity of the two-impurity model into a regular triangle such that the whole configuration turns into a straight triangular prism satisfying $C_2\otimes C_3$ symmetry. Each impurity of the rectangle model can also be extended into two impurities along the third direction to generate a cuboid of $N=8$, whose spectral function matrix $\rho_{\mu\nu}(\omega)$ can be diagonalized by the transformation matrix $w^{(2)}\otimes w^{(2)} \otimes w^{(2)}$. Thus, cuboid models are also solvable and the decomposition can be constructed exactly with $N_A=N=8$.

\begin{figure}[t]
	\begin{center}
		\includegraphics[width=8.6cm]{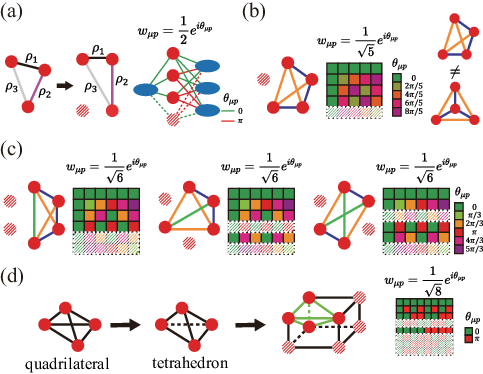}
	\end{center}
	\caption{(a) Illustration of the graph method by mapping a general triangle to the subgraph of a rectangle and then doing the decomposition by ignoring the additional impurity. (b)(c) 4-impurity configurations that can be embedded into a regular pentagon or hexagon. (b) shows two quadrilaterals with different graph structures (equivalent edges). (d) Mapping of a general quadrilateral to a tetrahedron and then to the subgraph of a cuboid, giving the decomposition with a reduced $N_A=8$ instead of $3\cdot2^{N-2} = 12$ from regular polygons.}
	\label{fig3}
\end{figure}

\paragraph{Graph method.}
For general configurations without (block) circulant symmetry, exact decomposition is also possible but requires a larger number of auxiliary baths ($N_A > N$). To prove this, we introduce a graph method and regard the real space configuration of a general impurity model as a weighted graph with certain topological (edge) structure. One important advantage of the graph method is that it allows for deformation of the configuration, while keeping the graph structure (equivalent edges) unchanged. The graph of the general model can then be mapped to the subgraph of a larger regular model (polygons or their extensions), whose exact solution can be used also to decompose the spectral function matrix of the original model. 

To give an example, we consider a general 3-impurity model whose real space configuration contains three different edges in correspondence with different off-diagonal elements in its spectral function matrix $\rho_{\mu\nu}(\omega)$. Since the decomposition only depends on the structure of the matrix rather than exact values of its elements, we can embed this graph into a rectangle whose 3-vertex subgraph also has three inequivalent edges as shown in Fig.~{\ref{fig3}}(a). 

The shared bath of the 3-impurity model can be decomposed using the same decomposition scheme for the rectangle model that satisfies $\rho^{(4)}_{\mu\nu}(\omega)=\sum_{p} w_{\mu p }w^{*}_{\nu p} \tilde{\rho}_{p}(\omega)$, where $\mu,\, \nu = 0,\, 1,\, 2$ denote the three impurities in the 3-vertex subgraph, $\mu,\,\nu=3$ refer to the last impurity, and $p=0,\cdots,3$ mark the four auxiliary baths. To see how it works, we first ignore the last impurity from the spectral function matrix ($\mu=3$ or $\nu=3$) and replace all remaining matrix elements with the corresponding ones from the 3-impurity model. Since this procedure does not affect the structure of the matrix, the same transformation matrix $w$ can also decompose the updated $3\times 3$ spectral function matrix:
\begin{equation}\label{eq7}
	\rho^{(3)}_{\mu\nu}(\omega) = \sum_{p} w_{\mu p}w^{*}_{\nu p} \tilde{\rho}_{p}(\omega),
\end{equation}
where $\mu,\,\nu=0,\,\cdots,\,N-1$ with $N=3$, and $p = 0,\,\cdots,\, N_A - 1$ with $N_A=4$. We have therefore an exact decomposition scheme of general 3-impurity models using a larger number of auxiliary baths, as illustrated in Fig.~\ref{fig3}(a) and proposed previously in Ref.~\cite{Hu2023}. This procedure can be readily extended to all impurity models and leads to the following statement:

\textbf{Theorem 3.} The spectral function matrix of any impurity configuration can be exactly decomposed if its corresponding graph can be embedded into a larger graph with (block) circulant symmetry.

More examples are given in Figs.~{\ref{fig3}}(b)(c), where four quadrilateral configurations ($N=4$) are strictly decomposed by embedding them into a regular pentagon ($N_A = 5$) or hexagon ($N_A=6$). For rectangle, this provides a different decomposition scheme with a larger $N_A=6$. Specifically, one may map any $N$-impurity configuration ($N> 2$) to the subgraph of a regular $N_A$-gon. We find that $N_A=3\cdot2^{N-2}$ can ensure the subgraph to have the largest number of inequivalent edges as in a general $N$-impurity model \cite{note}. Thus, exact decomposition of the shared bath is possible for any $N$-impurity model, albeit with $N_A>N$. On the other hand, one may choose other graphs for the embedding. All 3-impurity models can be embedded into a rectangle with $N_A=4$ instead of a regular hexagon with $N_A=6$, while a general 4-impurity model can be mapped to a tetrahedron with the same graph structure and then embedded into a cuboid with $N_A=8$ rather than $3\cdot2^{N-2}=12$ as shown in Fig.~\ref{fig3}(d). Thus, $3\cdot2^{N-2}$ is not the least number of the auxiliary baths. It remains an open question to find the minimum $N_A$ allowing for such embedding.

What then determines the decomposition? Although we start with highly symmetric models, it is actually the graph structure rather than the symmetry that plays a key role. A rhombus (diamond) has the same $D_2$ symmetry as the rectangle but cannot be decomposed with $N_A=N=4$, because their graphs contain very different equivalent edges and are not isomorphic \cite{Diestel2017}. Figure~\ref{fig3}(b) gives another example. A regular triangle with an additional point at the center differs topologically from an isosceles trapezoid because the former contains three equivalent edges sharing a common point. Although the triangle keeps $C_3$ symmetry, it cannot be decomposed with  $N_A=4$ or 5. One may, however, map the graph to another graph with less equivalent edges, namely, equivalent edges in the latter correspond also to equivalent edges in the former, but not vice versa, and then apply the transformation matrix of the latter to the former. In this way, the 4-impurity triangle may be deformed to a kite, the middle graph in Fig.~\ref{fig3}(c), and decomposed using the transformation matrix of a regular hexagon with $N_A=6$. We may regard the spectral function matrix $\rho_{\mu\nu}(\omega)$ as the adjacent matrix of a graph that measures the weight of each edge. Its $\omega$-dependency then introduces dynamics into the graph. This makes a connection with the graph theory and our approach corresponds to a mode decomposition of such dynamic graphs.

\begin{figure}[t]
	\begin{center}
		\includegraphics[width=8.6cm]{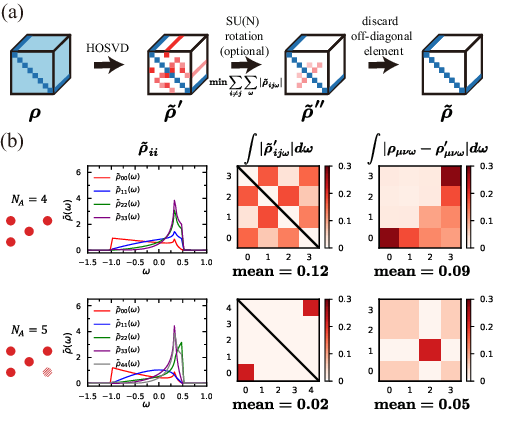}
	\end{center}
	\caption{(a) Illustration of the approximate decomposition procedure using the HOSVD. (b) Comparison of approximate solutions with $N_A=4$ and 5 for a special 4-impurity model where three impurities form a right-angled triangle and the rest one is located at the midpoint of the hypotenuse. Also compared are the errors and their means induced by discarding the off-diagonal elements after the HOSVD.}
	\label{fig4}
\end{figure}

\paragraph{Approximate decomposition.}
Although exact decomposition may be applied to all configurations, the number of auxiliary baths may be too large to perform realistic NRG calculations for irregular models. It is therefore important to find approximate algorithms that can reduce the computational workload. Here we propose a numerical procedure using the high-order singular value decomposition (HOSVD) \cite{Kolda2009}. 

We first discretize $\omega$ and treat the spectral function matrix $\rho_{\mu \nu}(\omega)$ as a three-order tensor $\rho_{\mu\nu\omega}$. The HOSVD algorithm may then be applied and yields
\begin{equation}\label{eq4}
	\rho_{\mu\nu\omega} = \sum_{ij}w_{\mu i} w^{*}_{\nu j} \tilde{\rho}^\prime_{ij\omega},
\end{equation}
where $\tilde{\rho}^\prime_{ij\omega}$ is typically non-diagonal with respect to $i$ and $j$. As illustrated in Fig.~\ref{fig4}(a), we further discard all off-diagonal elements after an optional rotation to minimize their $\ell_1$ norm and reduce the subsequent error. The auxiliary baths are given by $\tilde{\rho}_{p\omega}=\tilde{\rho}^{\prime\prime}_{pp\omega}$. The error introduced by discarding the off-diagonal elements can be measured by  
\begin{equation}\label{eq9}
	{\rm{error}} = \frac{1}{N^2}\sum_{\mu\nu}\int |\rho_{\mu \nu \omega}-\rho'_{\mu \nu \omega}| d\omega,
\end{equation} 
where $\rho'_{\mu\nu\omega}$ are calculated from $\tilde{\rho}_{p\omega}$ using Eq.~(\ref{basic}). This gives an approximate decomposition with $N_A=N$ auxiliary baths for any $N$-impurity model. 

A decomposition with larger $N_A$ may be constructed simply by adding $N_A - N$ fictitious impurities to the initial configuration and then eliminating them after the decomposition. An example is given in Fig.~\ref{fig4}(b), where two approximate solutions of a special 4-impurity model are constructed with $N_A = 4$ and 5 auxiliary baths, respectively. The model contains three impurities forming a right-angled triangle and one additional impurity at the midpoint of the hypotenuse. The off-diagonal elements of $\tilde{\rho}^\prime_{ij\omega}$ and the errors introduced by discarding them are shown for comparison. We find only about 10\% mean errors for $N_A=4$, which already allows for a qualitative study of the original model. For $N_A=5$, the errors are further reduced, possibly because the extended 5-impurity model has a higher level of symmetry. In general, a better solution may be expected with a larger $N_A$ by optimizing the positions of the fictitious impurities.

\begin{table*}[t]
\caption{\label{tab} Collection of some simple exact decomposition schemes.}
\centering
\begin{ruledtabular}
\begin{tabular}{cccccc}
Configuration & $N$ & $N_A$ & spectral function matrix & symmetry group & transformation matrix \\
\hline
line & 2 & 2 & \multirow{5}{*}{circulant} &  $C_2$ & $w^{(2)}$ \\
regular triangle & 3 & 3 &  &  $C_3$ & $w^{(3)}$ \\
square & 4 & 4 &  &  $C_4$ & $w^{(4)}$ \\
regular hexagon & 6 & 6 & &  $C_6$ & $w^{(6)}$ \\
regular $n$-gon  & $n$ & $n$ &&  $C_n$ & $w^{(n)}$ \\
\hline
rectangle & 4 & 4 & \multirow{5}{*}{block circulant} &  $C_2 \otimes C_2$ & $w^{(2)} \otimes w^{(2)}$ \\
straight triangular prism & 6 & 6 & &  $C_2 \otimes C_3$ & $w^{(2)} \otimes w^{(3)}$ \\
cuboid & 8 & 8 &  &  $C_2 \otimes C_2 \otimes C_2$ & $w^{(2)} \otimes w^{(2)}  \otimes w^{(2)}$ \\
\multirow{2}{*}{$d$-dim cuboid} & \multirow{2}{*}{$2^d$} & \multirow{2}{*}{$2^d$} & & {$\underbrace{C_2 \otimes C_2 \otimes \cdots \otimes C_2}$} & {$\underbrace{w^{(2)} \otimes w^{(2)} \otimes \cdots \otimes w^{(2)}} $} \\
& & & & $d$ & $d$\\
\hline
\multirow{2}{*}{trisosceles trapezoid} & \multirow{2}{*}{4} & \multirow{2}{*}{5, 6} & submatrix of regular pentagon  &  \multirow{2}{*}{-} & \multirow{2}{*}{submatrix of $w^{(5)}$ or $w^{(6)}$} \\
& & & or regular hexagon & & \\
kite & 4 & 6 & submatrix of regular hexagon & - & submatrix of $w^{(6)}$ \\
\hline
general 3-impurity models & 3 & $\le4$ & submatrix of rectangle &  - &  submatrix of $w^{(2)} \otimes w^{(2)}$ \\
general 4-impurity models & 4 & $\le8$ & submatrix of cuboid &  - & submatrix of $w^{(2)} \otimes w^{(2)} \otimes w^{(2)}$\\
general 5-impurity models & 5 & \multirow{2}{*}{$\le16$} & \multirow{2}{*}{submatrix of 4-dim cuboid} &  \multirow{2}{*}{-} & submatrix of\\
general 6-impurity models & 6 & & &  & $w^{(2)} \otimes w^{(2)} \otimes w^{(2)} \otimes w^{(2)}$ \\
\end{tabular}
\end{ruledtabular}
\end{table*}

\paragraph{Discussion.} 
Last, we comment on the non-negativity of $\tilde{\rho}_{p}(\omega)$. This is easy to prove for regular impurity configurations with $N_A=N$ since the $N\times N$ transformation matrix $w_{\mu p}$ is invertible and $\tilde{\rho}_{p}(\omega)$ may be expanded according to the eigenstates (momentum) of the original noninteracting bath. The proof can be readily extended to more general cases where the $N$-impurity configuration is a subset of a regular $N_A$-impurity configuration with invertible $N_A\times N_A$ transformation matrix. Negative $\tilde{\rho}_{p}(\omega)$ may only occur if the impurity configuration is deformed, embedded, and decomposed based on another configuration, because the bath's Hamiltonians are not preserved during the embedding and $\tilde{\rho}_{p}(\omega)$ can no longer be expanded using their eigenstates. One example is the 3-impurity model with $N_A=4$ based on rectangle, where one of $\tilde{\rho}_{p}(\omega)$ becomes negative near the upper band edge \cite{Hu2023}. In this case, one may set the negative values to zero, which introduces additional errors but may still be used if it is not close to the Fermi energy. The negative values imply that the shared bath induces entanglements between distant impurities that cannot be fully captured by the chosen mapping scheme. Similarly, for approximate decomposition using HOSVD, the non-negativity is guaranteed with invertible $w_{\mu p}$ for any $N_A$ as long as the $N_A-N$ fictitious impurities to generate the decomposition are all chosen from real lattice sites of the original bath. Negative $\tilde{\rho}_{p}(\omega)$ may occur only for artificial $\rho_{\mu\nu}(\omega)$ elements assigned to the fictitious impurities.

For convenient usage, we collect in Table \ref{tab} some of the exact decomposition schemes for simple impurity configurations, including regular ones with (block) circulant symmetry and irregular ones that can be embedded into a larger regular graph. This provides a systematic strategy for solving the most interesting quantum impurity models using the auxiliary-bath NRG approach. Our results are not limited to free electron baths but can be applied to all baths with a bilinear Hamiltonian that can be integrated out to yield a spectral function matrix of similar forms. We emphasize again that it is the graph structure that plays a more fundamental role in the decomposition. Non-isomorphic graphs with the same symmetry group may not have the same decomposition property. This connection with the graph theory may help design more efficient algorithm for NRG calculations.

This work was supported by the National Natural Science Foundation of China (NSFC Grant No. 12174429), the Postdoctoral Fellowship Program of CPSF (Grant No. GZC20232943), the National Key Research and Development Program of China (Grant No. 2022YFA1402203), and the Strategic Priority Research Program of the Chinese Academy of Sciences (Grant No. XDB33010100).

J.-B. W. and D. H. contributed equally to this work.


\begin{thebibliography}{99}
\bibitem{note} This can be proved by mapping the $\mu$-th impurity ($\mu=0,\cdots,N-1$) to the $i_\mu$-th point of the regular $N_A$-gon ($i_\mu\in\{0,\cdots,N_A-1\}$). To guarantee that the subgraph contains no equivalent edges, we may require $i_{\mu}-i_{\mu-1} = i_{\mu-1} -i_0+1$ and $i_0=0$, such that the distance between impurities $\mu$ and $\mu-1$ is larger than those of all impurities before them. This gives $i_\mu=2^{\mu}-1$. The minimum number of auxiliary baths is then $N_A=i_{N-1}+1+i_{N-1}-i_{N-2}=3\cdot2^{N-2}$ to ensure that all distances ($N_A-i_\mu+i_\nu$) remain different if counted from opposite direction of the regular $N_A$-gon.
\end{thebibliography}
\end{document}